\documentclass{iaus}   
\usepackage{graphicx}

\def\HST{{\it HST}}  
\def\deg{\ifmmode ^{\rm o} \else $^{\rm o}$\fi}

\def\kms{\ifmmode {\rm km\,s}^{-1} \else km\,s$^{-1}$\fi}  
\def\micron{\ifmmode \mu{\rm m} \else $\mu$m\fi}  
\def\Msun{\ifmmode {\rm M}_{\odot} \else M$_{\odot}$\fi}  
\def\Lsun{\ifmmode {\rm L}_{\odot} \else L$_{\odot}$\fi}

\def\vFWHM{\ifmmode v_{\mbox{\tiny FWHM}} \else  
            $v_{\mbox{\tiny FWHM}}$\fi}  
\def\CCF{\ifmmode F_{\it CCF} \else $F_{\it CCF}$\fi}  
\def\ACF{\ifmmode F_{\it ACF} \else $F_{\it ACF}$\fi}  
\def\Halpha{\ifmmode {\rm H}\alpha \else H$\alpha$\fi}  
\def\Hbeta{\ifmmode {\rm H}\beta \else H$\beta$\fi}  
\def\Hgamma{\ifmmode {\rm H}\gamma \else H$\gamma$\fi}  
\def\Hdelta{\ifmmode {\rm H}\delta \else H$\delta$\fi}  
\def\Lya{\ifmmode {\rm Ly}\alpha \else Ly$\alpha$\fi}  
\def\Lyb{\ifmmode {\rm Ly}\beta \else Ly$\beta$\fi}

\def\ciii{\ifmmode {\rm C}\,{\sc iii} \else C\,{\sc iii}\fi}  
\def\civ{\ifmmode {\rm C}\,{\sc iv} \else C\,{\sc iv}\fi}

\def\o5007{[O\,{\sc iii}]\,$\lambda5007$}

\def\Msigma{$M_{\rm BH}$--$\sigma_*$}  
\def\MLbulge{$M_{\rm BH}$--$L_{\rm bulge}$}  
   
\title[Central Black Hole Masses] 
{Toward Precision Measurement of Central Black Hole Masses}

\author[B.M.\ Peterson]   
{Bradley M.\ Peterson$^1$}   
   
\affiliation{$^1$Department of Astronomy and Center for    
Cosmology and AstroParticle Physics, The Ohio State University,   
140 West 18th Avenue, Columbus, OH 43210, USA\\   
Email: {\tt peterson@astronomy.ohio-state.edu}}   
   
\pubyear{2010}   
\volume{267}  
\pagerange{119--126}   
\jname{Co-Evolution of Central Black Holes and Galaxies}   
\editors{B.M.\ Peterson, R.S.\ Somerville, \& T.\ Storchi-Bergmann, eds.}   
\begin{document}   
   
\maketitle   
   
\begin{abstract}   
We review briefly direct and indirect methods of measuring the  
masses of black holes in galactic nuclei, and then focus  
attention on supermassive black holes in active nuclei,  
with special attention to results from reverberation mapping  
and their limitations. We find that the intrinsic scatter in
the relationship between the AGN luminosity and the broad-line
region size is very small, $\sim\! 0.11$\,dex, comparable to
the uncertainties in the better reverberation measurements. We also
find that the relationship between reverberation-based
black hole masses and host-galaxy bulge luminosities
also seems to have surprisingly little intrinsic scatter,
$\sim\! 0.17$\,dex. We note, however, that there are still
potential systematics that could affect the overall mass
calibration at the level of a factor of a few.   
  
\keywords{galaxies: active, galaxies: nuclei, techniques: spectroscopic}   
  
\end{abstract}   
   
\firstsection 
\section{Introduction}   
With the advent of suitable technology for high angular resolution    
spectroscopy, from space with {\em Hubble Space Telescope} and    
from the ground with adaptive optics assistance, it has become    
possible to measure the masses of the central black holes in    
many nearby galaxies. Observations of nearby galaxies have    
led to the identification of scaling relationships that then allow    
us to estimate masses of black holes in distant galaxies and    
thus determine, at least in principle, the mass function for    
supermassive black holes, both locally and over cosmic time.    
Consequently, there have been tremendous advances in our    
understanding of the evolution of the supermassive black hole    
population and that of the galaxies that host them over the    
history of the universe. Despite this incredible progress, it is    
important to understand that supermassive black hole    
measurement is a field in its infancy: even the methods deemed    
most reliable have systematic uncertainties that have not been    
mitigated in a completely satisfactory way. Masses based on    
stellar dynamics and gas dynamics are highly model dependent    
and are only as reliable as the underlying assumptions. An    
obvious example is the recent work of   
\cite[Gebhardt \& Thomas (2009)]{GebThom09}  
who found that the mass of 
the black hole in M\,87, widely regarded as    
one of the most secure measurements, changed by a factor of    
two with the introduction of a dark matter halo into the models.    
It is also worth noting that only two black hole measurements,
the Milky Way (based on orbital motion of stars in    
the Galactic Center) and NGC 4258 (based on motions of    
megamasers), meet the formal criterion of measuring the mass    
on a sufficiently compact scale that the black-hole nature of the    
central source is proven. In all other present measurements, it    
remains an article of faith that the central source is a    
singularity, although it is hard to see practically how these    
masses could be anything else. We will therefore adhere to 
the convention of referring to these objects as black holes.   
   
Fundamentally, mass measurements are based on how the    
central black holes accelerate nearby matter, either stars or gas.    
The advantage of using stellar dynamics is that stars respond    
only to gravitational forces. The corresponding disadvantage is    
that high angular resolution is necessary; measurement of    
stellar motions at large distances from the nucleus makes the    
mass determinations more uncertain. Gas motions, on the other    
hand, allow us to probe closer to the nucleus; indeed, in the    
case of reverberation mapping, angular resolution is irrelevant    
since time resolution substitutes for it. The disadvantage of    
using gas motions is, of course, that gas also responds to forces    
other than gravity, with radiation pressure being the main    
concern (see \S\ref{section:radpress}).    
   
It is important to distinguish between {\em direct} and {\em    
indirect} methods of mass measurement. {\em Direct    
methods} are based on dynamics of gas or stars accelerated by    
the central black hole. This would include stellar dynamics, gas    
dynamics, and reverberation mapping. {\em Indirect methods},    
on the other hand, are based on observables correlated with the    
mass of the central black hole. Indirect methods include the    
\Msigma\ and \MLbulge\ relationships, the fundamental plane,    
as well as AGN scaling relationships such as the BLR   
radius-luminosity relationship that we will discuss later.

We also sometimes refer to ``primary,'' ``secondary,'' and    
``tertiary'' methods, with the difference  
based on model-dependence and assumptions required. Primary methods    
require fewer assumptions and have little model dependence:    
the masses of the black holes in the Milky Way and NGC    
4258, based on proper motions and radial velocities, are thus    
certainly primary. Stellar dynamics and gas dynamics are    
sometimes also regarded as primary methods as they do not    
hinge on other results. On the other hand, reverberation    
mapping, as it has been practiced to date, currently depends on    
other ``primary direct'' methods for a zero point for the mass    
scale, so it is technically a ``secondary method,'' though    
it is a still a ``direct method.''   
  
\begin{figure}[b]  
\begin{center}  
 \includegraphics[width=5.2in]{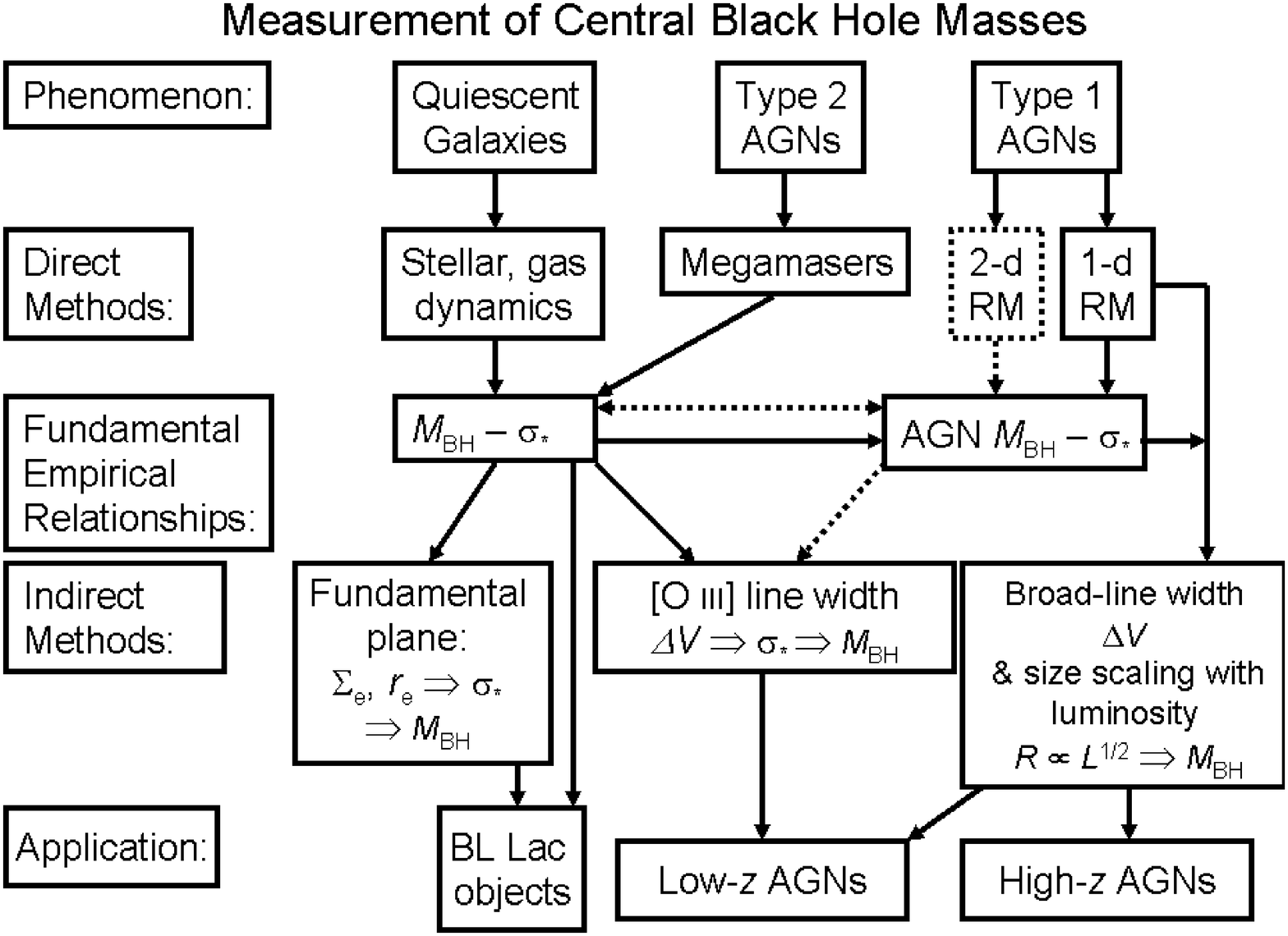}   
 \caption{Methods and applications of black hole mass measurements  
in galactic nuclei.}  
   \label{fig1}  
\end{center}  
\end{figure}  
  
Figure 1 illustrates in flowchart format how masses are    
determined for both quiescent and active galaxies. It is slightly    
oversimplified in that stellar and gas dynamics have been used    
to measure central masses of a very limited number of AGNs,    
as discussed below. One-dimensional (1-d)  
reverberation-mapping refers to measurement of mean emission-line    
response times, i.e., reverberation mapping as it has been    
practiced to date. As noted above, this requires some external    
calibration of the mass scale and this is currently provided by    
assuming that the \Msigma\ relationship is the same in    
quiescent and active galaxies. With two-dimensional (2-d)    
reverberation mapping, we aim to produce a velocity--delay    
map which should give us sufficient information to model the    
BLR dynamics and determine the central masses;   
model-dependence will still be something of an issue, as it is with    
stellar and gas dynamics. But this will eventually put    
reverberation mapping on a par with stellar and gas dynamics    
as a ``primary direct'' method and it will be possible to    
compare directly the \Msigma\ relationships in quiescent and    
active galaxies rather than assume that they are the same. This    
may occur in the very near future: the first reliable detections    
of velocity-dependent lags were reported at this conference    
(see the papers by Bentz and Denney, these proceedings).   
   
In Table 1, we list published masses of a few relatively nearby    
AGNs for which black hole measurements have been made    
using more than one method. The most accurate measurement    
is provided by megamaser motions in NGC 4258. However,    
megamaser sources like this are exceedingly rare. Moreover,    
megamasers occur in type 2 AGNs where our view of the    
central engine is obscured, thus making reverberation mapping    
impossible. Mass measurements by stellar dynamics and/or gas    
dynamics have been made for a handful of nearby AGNs (e.g.,    
Hicks et al., these proceedings), but in both cases, significant future    
progress depends on dramatic improvements in angular    
resolution. Thus, if only by default, reverberation mapping is    
the near-term future for AGN black hole mass    
measurement. The downside to this is that reverberation    
mapping is a resource-intensive technique  
based on high-precision spectrophotometry. However, 
as we see below, reverberation results already provide us 
with a shortcut to mass estimation.  
   
\begin{table}   
  \begin{center}   
  \caption{Comparison of Black Hole Mass Measurements}   
  \label{tab1}   
  \begin{tabular}{lcccc}\hline    
{\bf Method} & {\bf NGC 4258} & {\bf NGC 3227} & {\bf NGC 4151} \\    
& \multicolumn{3}{c}{(Units $10^6\,M_\odot$)}\\ \hline   
\underline{Direct methods:} \\   
Megamasers & $38.2 \pm 0.1^{[1]}$ & N/A & N/A \\   
Stellar dynamics & $33 \pm 2^{[2]}$ & 7--20$^{[3]}$ & $\leq 70^{[4]}$ \\   
Gas dynamics & 25--260$^{[5]}$ & $20^{+10}_{-4}$ $^{[6]}$ & $30^{+7.5}_{-22}$ $^{[6]}$ \\   
Reverberation & N/A & $7.63^{+1.62}_{-1.72}$ $^{[7]}$ & $46\pm5^{[8]}$ \\   
\underline{Indirect methods:}\\   
$M_{\rm BH}$--$\sigma_{*}^{[9]}$ & 13 & 25 & 6.1 \\   
$R$--$L$ scaling$^{[10]}$ & N/A & 15 & 29--120 \\   
\hline   
  \end{tabular}   
\end{center}   
   
  $^{[1]}$Herrnstein et al.\ (2005).   
  $^{[2]}$Siopis et al.\ (2009).   
  $^{[3]}$Davies et al.\ (2006).   
  $^{[4]}$Onken et al.\ (2007).   
  $^{[5]}$Pastorini et al.\ (2007).    
  $^{[6]}$Hicks \& Malkan (2008).   
  $^{[7]}$Denney et al.\ (2010).   
  $^{[8]}$Bentz et al.\ (2006b).   
  $^{[9]}$G\"{u}ltekin et al.\ (2008).   
  $^{[10]}$McGill et al.\ (2008).   
\end{table}

\section{Reverberation-Based Mass Measurements}   
   
Reverberation mapping 
(\cite[Blandford \& McKee 1982]{BMcK82};
\cite[Peterson 1993]{Pet93})
makes use of the time-delayed    
response of the broad emission lines to continuum variations to    
provide a size estimate for the broad-line region (see   
Peterson 2001  
for a fairly thorough tutorial). At the present time,    
emission-line time delays, or lags $\tau$, have been measured    
for $\sim45$ AGNs, in nearly all cases for \Hbeta, but in many    
cases for the other Balmer lines and in a few cases for multiple    
lines extending into the ultraviolet. We can then obtain the    
mass of the black hole (or, more accurately, the mass enclosed    
out to the distance $R =c\tau$) by combining this with the    
emission-line width $\Delta V$, i.e.,   
\begin{equation}   
\label{equation:virial}
M_{\rm BH} =   \frac {f \left( \Delta V^2 R \right)}{G},    
\end{equation}   
where $f$ is a dimensionless factor of order unity that depends    
on the structure, dynamics, and orientation of the BLR. If, for    
example, the BLR is a thin ring of material in a Keplerian    
orbit of inclination $i$ around the black hole, then   
$f \propto 1/\sin^2 i$. Of course, the real geometry of the BLR must    
be much more complex than a simple ring or disk; unified    
models suggest that we see type 1 AGNs at inclinations $0\deg    
\lesssim i \lesssim 45\deg$, in which case the observed line-of-sight    
velocities in a ring or disk are a small projection of the orbital    
speed. Indeed, for $i\lesssim 20\deg$, the projected line-of-sight 
velocity width of the line is so small that  
we would underestimate the mass of the black hole by more than 
an order of magnitude. We note in passing that    
emission-line lags are unaffected by inclination as long as the    
system has axial symmetry and the line emission is isotropic.    
Nevertheless, there is evidence, mostly from radio-loud    
systems, that inclination is an important element in the    
determination of $f$. For example, the ratio of core-to-lobe    
power in double radio sources correlates inversely with line    
width   
\cite[(Wills \& Browne 1986)]{Wills1986};   
core-dominated systems,    
where we are looking down the system axis, have narrower lines    
than systems at larger inclination. Similarly, flat-spectrum    
radio sources, in which again we are observing close to the axis    
of the system, have narrower emission lines than steep    
spectrum sources, which are seen at higher inclination  
\cite[(Jarvis \& McLure 2006)]{Jarvis2006}.  
But the differences are not extraordinarily    
large: for example, Jarvis \& McLure find that for their sample    
of radio-loud AGNs, sources with $\alpha > 0.5$ have a mean    
FWHM of 6464\,\kms\ for \Hbeta\ and Mg\,{\sc ii}, while    
those with $\alpha < 0.5$ have a mean width of 4990\,\kms.    
These results clearly indicate that (1) inclination is important,    
but (2) the BLR gas has considerable velocity dispersion in the polar    
direction. Given this, we now have to recognize that a direct    
comparison between reverberation-based masses and, say,    
stellar or gas dynamics, will depend very much on the    
generally unknown inclination of the BLR. We should generally 
not expect good agreement between reverberation masses 
and masses measured by other techniques for individual
galaxies unless the  
inclination is known or strongly constrained. 
   
On the other hand, given a sample of AGNs, we can determine    
an {\em average} value $\langle f \rangle$ by normalizing the    
AGN \Msigma\ relationship to that of quiescent galaxies. A    
first attempt to do this yielded a value $\langle f \rangle = 5.5    
\pm 1.8$ 
\cite[(Onken et al.\ 2004)]{Onken04}: 
while this intuitively seems a little high, we can    
again plausibly attribute this to inclination 
effects and our predisposition     
the inclinations of type 1 AGNs are typically rather    
small. The main liability of this particular calibration  
of the mass scale is that it 
is based on low-redshift, low-luminosity objects.    
All of the $\sigma_*$ measurements are based on the Ca\,{\sc    
ii} triplet which is unfortunately redshifted into  
atmospheric water vapor    
bands at $z > 0.06$. We are now attempting to measure    
$\sigma_*$ at higher redshift by using observations of the CO    
bandhead in the {\em H}-band (1.6\,\micron), a process greatly    
aided by use of adaptive optics with integral field units that   
concentrate the bright AGN into the central few pixels  
and at the same time integrate a few arcseconds in each 
coordinate without loss of spectral resolution (see the
papers by Dasyra and Grier in    
these proceedings). In addition to new results at higher    
redshifts and luminosities, the LAMP collaboration (Bentz,    
these proceedings) is adding additional  masses and velocity    
dispersions at the low-mass end of the AGN distribution. As a    
consequence of expanding the range in black hole mass, the    
slope of the AGN \Msigma\ relationship is becoming better    
defined and shows improved consistency with that for    
quiescent galaxies (Woo, these proceedings).    
   
If we assume that the AGN \Msigma\ relationship has the    
same zero point and slope as that for quiescent galaxies, a    
maximum likelihood analysis places an upper limit on the    
intrinsic scatter in the relationship of $\Delta \log M_{\rm BH}    
\approx 0.40$\,dex, which is consistent with what is found for    
quiescent galaxies (G\"{u}ltekin, these proceedings). This is a    
reassuring indication that the masses derived from    
reverberation mapping are reliable. There are, of course, other    
such indications, including the direct comparisons of a few    
cases as in Table 1. Also, in each case in which lags have been    
measured for multiple emission lines in a single source, there is    
an anticorrelation between lag and line width that is consistent    
with a constant virial product $\Delta V^2 R$  
\cite[(Peterson \& Wandel 1999)]{Peterson1999}.  
Finally, AGNs show the same correlation    
between black hole mass and bulge luminosity   
(the \MLbulge\ relationship) that is seen in quiescent galaxies    
\cite[(Bentz et al.\ 2009b)]{Bentz2009b}.   
Indeed,  a maximum likelihood analysis    
gives an upper limit to intrinsic scatter,  $\Delta \log M_{\rm    
BH} \approx  0.17$\,dex, which is actually smaller than the    
intrinsic scatter in this relationship for quiescent galaxies    
($\Delta \log M_{\rm BH} \approx  0.38$\,dex;   
\cite[G\"{u}ltekin et al.\ 2008]{Gultekin2008}).   
   
\section{\boldmath BLR Scaling with Luminosity: The $R$--$L$ Relationship}   
   
The emission-line spectrum from an ionized gas, setting aside    
elemental abundances, is largely controlled by the particle
density $n$ and an ionization parameter   
\begin{equation}   
U = \frac{Q({\rm H})}{4 \pi R^2 n c},   
\end{equation}   
where $Q({\rm H})$ is the number of hydrogen-ionizing    
photons emitted per second by the central object. To some low    
order of approximation (principally ignoring the Baldwin    
Effect), all AGN spectra have similar emission-line flux 
ratios and similar emission-line equivalent widths,    
suggesting that $U$ and $n$ do not vary much from one source    
to another. If we further assume that $Q({\rm H}) \propto L$    
where $L$ is the luminosity of the central source in some    
arbitrary band, we are led to the na\"{\i}ve prediction that $R    
\propto L^{1/2}.$   
   
Thus since the early days of photoionization equilibrium    
calculations, some such relationship between the BLR radius    
and the continuum luminosity (hereafter simply the $R$--$L$    
relationship) of the power-law form $R \propto L^{\alpha}$    
had been anticipated. Observationally the $R$--$L$    
relationship for \Hbeta\ first emerged as statistically significant    
with a slope $\alpha \approx 0.7$ with the addition of 17    
PG QSOs by   
\cite[Kaspi et al.\ (2000)]{Kaspi00}  
to the existing similar-sized    
reverberation-mapping database on Seyfert 1 galaxies. It is    
sometimes stated that this result was enabled by extension of the    
luminosity range by nearly two orders of magnitude; it is more    
correct to say, however, that it was due to the fact that the    
QSOs are so much more luminous than their host galaxies that    
starlight contributes negligibly to the luminosity measurement.     
Indeed, it is surprising how much starlight contaminates AGN    
luminosity measurements in the optical, though this problem is  
certainly exacerbated in reverberation monitoring data as large    
spectrograph apertures and spectral extraction windows are    
used to mitigate the effects of variable seeing on the total flux    
measurements. Bentz et al.\ (2006a, 2009a)
have attempted to    
account for starlight contamination of optical luminosities of    
AGNs by using unsaturated high-resolution optical images    
obtained with \HST/ACS. The surface brightness distribution is    
modeled so that the AGN point source can be removed and the    
galaxy contamination estimated by simulated aperture    
photometry. When this is done, the slope of the $R$--$L$    
relationship for \Hbeta\ is found to be $\alpha \approx 0.5$,    
remarkably consistent with the na\"{\i}ve prediction.   
   
This leads us to the remarkable realization that it is possible to    
estimate masses of the black holes in AGNs from only a single    
spectrum: by measuring the luminosity, we infer the BLR    
radius and we can combine this with the emission-line width to    
measure the mass 
\cite[(Wandel, Peterson, \& Malkan 1999)]{Wandel1999}. 
This is variously referred to as the    
``photoionization method,'' ``single-epoch spectrum method,''  
or simply ``masses from scaling relationships.''  
   
An interesting question at this point is how much intrinsic    
scatter is in the $R$--$L$ relationship? This is of keen interest for    
two reasons: first, this will set a fundamental limit on the    
accuracy with which we can estimate masses from single    
spectrum, and second, it dictates future observing strategies for    
further refinement of the $R$--$L$ relationship. If the intrinsic    
scatter is large, then many more reverberation measurements    
will be needed to beat down the statistical noise in the    
relationship. On the other hand, if the scatter is small, then    
improvements in the determination of the $R$--$L$    
relationship will come from obtaining {\em better}    
reverberation data on a smaller number of objects.   
   
\begin{figure}[b]  
\begin{center}  
 \includegraphics[width=4.5in]{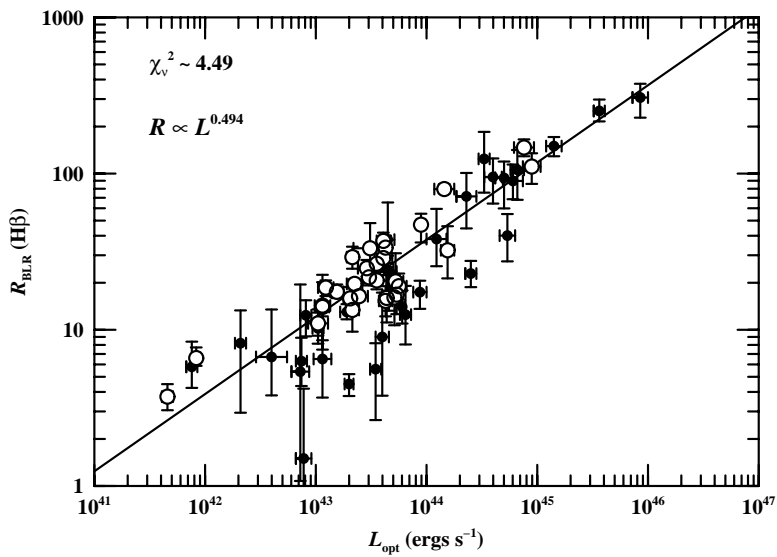}   
 \caption{The $R$--$L$ relationship for H$\beta$.
The luminosity is $\lambda L_\lambda (5100\,{\rm \AA})$
and the BLR radius is measured in light days.
The open circles indicate the highest-quality
measurements. The slope of the fit to the highest-quality
data is indistinguishable from that to all the
data, but the scatter is only 0.11 dex.}
   \label{fig2}  
\end{center}  
\end{figure}

To proceed with this, we want to start by using only the 
very best data sets, since these are ones where intrinsic 
scatter is most easily isolated: there remain elements of 
both art and luck in reverberation studies, and the ugly 
truth is that not all of the light curves give as clean 
a result as we would like. To minimize the impact of the 
lower-quality data, Catherine Grier and I independently visually 
inspected all available light curves (for the optical    
continuum and \Hbeta\ emission line only) with the intent of    
identifying the ``best'' reverberation data: we used a very    
simple criterion, that you could easily see the same pattern of    
variability in the continuum and \Hbeta\ light curves 
and could estimate the   
lag simply by inspection
(see Figure \ref{fig2}). About  half of the existing light    
curves met this criterion. A maximum likelihood analysis of    
these data indicates that the intrinsic scatter amounts to $\Delta    
\log R \approx 0.11$\,dex. This is really quite remarkable when    
one considers that the typical formal errors on this subset of the    
best reverberation data amount to  $\Delta \log R \approx    
0.09$\,dex. From this, we conclude that, at least over the    
luminosity and redshift range where it has been calibrated    
($41.5 \lesssim  \log \lambda L_{\lambda}  ({\rm erg\ s}^{-1})  \lesssim    
45$ at $\lambda = 5100$\,\AA\ and $z \approx 0$), the $R$--$L$    
relationship is as {\em statistically}  
effective as reverberation for obtaining BLR    
sizes and central black hole masses.  If you are wondering 
why we emphasize the ``statistical'' aspect of $R$--$L$ scaling, 
we refer you to the results of the ``indirect methods'' in 
Table 1: the results for individual sources can be quite misleading, 
as they certainly are in the case of NGC 4151, which
is a notorious outlier in the \Msigma\ relationship.

\section{\boldmath The $R$--$L$ Relationship at High    
Redshift and Indirect Mass Measurements}   

\subsection{The $R$--$L$ Relationship for UV Lines}
With the $R$--$L$ relationship, we are able to explore the    
black hole mass function, not only locally but at high redshift,    
enabling us to trace the history of black hole growth. Some    
exploratory work has been done on this and in fact there are    
claims that the \Msigma\ and \MLbulge\    
relationships evolve over time (e.g., Woo, these proceedings), 
although at least 
some published claims of evolution of the \Msigma\ 
relationship are clearly attributable to Malmquist bias. 
A lot of additional careful work will be required to sort this out.   
   
An obvious problem with estimating black hole masses at high    
redshift, of course, is that the \Hbeta\ emission line is    
redshifted out of the visible window at only modest redshift.    
We are thus forced to use other emission lines for which    
reverberation measurements are actually quite scarce: of the    
other potential broad lines for mass measurement, there is but a    
single reliable measurement of a Mg\,{\sc ii} lag   
\cite[(Metzroth, Onken, \& Peterson 2006)]{Metzroth2006}  
and a bare handful of C\,{\sc iv}    
lags, though these span quite a large range in luminosity,    
$39.5 \lesssim  \log \lambda L_{\lambda}({\rm erg\ s}^{-1})  \lesssim    
47$  at $\lambda = 1350$\,\AA\ in the rest frame  
(Kaspi et al.\ 2007 and references therein).   
One of these objects is at $z=2.17$, and    
the rest are at $z < 0.06$.    

The C\,{\sc iv}\,$\lambda1549$ emission line is especially    
important as it allows us to reach quite large redshifts in the    
optical.   
\cite{Vestergaard2002}  
first used C\,{\sc iv} to show that    
quasars with black hole masses of $\sim10^9\,M_\odot$ were    
already assembled by $z \sim 4$. She assumed that the    
C\,{\sc iv} $R$--$L$ relationship has the same slope as    
$\Hbeta$ and calibrated the C\,{\sc iv}-based mass scale by    
using single-epoch spectra of reverberation-mapped AGNs.    

\subsection{Characterizing the Velocity Field}
If there is an ``Achilles' heel'' in determining masses
based on individual spectra, it is probably {\em not}
the $R$--$L$ relationship, but rather how we characterize the
velocity field of the BLR. The line profile is
expected to be sensitive to both inclination and
Eddington rate (e.g., \cite[Collin et al. 2006]{Collin2006}).

In reverberation mapping
experiments, we obtain the best results by measuring
the line width in the ``rms spectrum'' formed by
computing the pixel-by-pixel variance in
all of the spectra  obtained in the reverberation
experiment. This isolates the variable part of the
emission line, which arises in the very gas for which
we are measuring the time delay. There are two common
ways to characterize the line width, either FWHM or
the line dispersion $\sigma_{\ell}$, which is the second 
moment of the line. The latter seems to give somewhat
better results (\cite[Collin et al. 2006]{Collin06};
\cite[Peterson et al. 2004]{Peterson04}). 
For either line-width measure, the modest
scatter, e.g., in the \MLbulge\ relationship, suggests
that the mass measurements are not particularly sensitive
to how we characterize the line width and thus
the velocity field of the BLR.

The situation is rather different with single-epoch
spectra where contamination by other features becomes
a serious issue. \cite{Denney2009} have explored
this in some detail for H$\beta$ by seeing how well
single spectra from reverberation campaigns reproduce
the actual reverberation-based masses. They find some
obvious effects (e.g., sensitivity of FWHM to proper
removal of the narrow-line component) and some
more subtle effects (e.g., $\sigma_\ell$ is badly affected
by blending, particularly when the AGN continuum is
in a faint state and the lines are especially broad).

Arguments about whether particular UV lines can be used
to estimate masses are really about how well the
widths of different lines in AGN spectra are correlated.
In the case of Mg\,{\sc ii} $\lambda2798$, the width of the    
Mg\,{\sc ii} line appears to be a good surrogate for the width of    
\Hbeta\ (Woo, these proceedings). 
\cite[Shen et al.\ (2008)]{Shen2008}, using a large collection of    
SDSS spectra, find that $\log \left(\left[{\rm    
FWHM}(\Hbeta)\right]/   
\left[ {\rm FWHM}({\rm Mg}\,\mbox{\sc ii})\right]\right) = 0.0062$, with    
scatter of only $\sim 0.11$\,dex.   
\cite[Onken \& Kollmeier (2008)]{Onken2008}    
find that the masses derived from \Hbeta\ and    
Mg\,{\sc ii} differ as a function of Eddington ratio, although    
they argue that it is possible to correct for this bias.   
   
The accuracy that can be obtained using the C\,{\sc iv}    
line remains somewhat controversial (see the contributions 
to these proceedings  by Vestergaard, Netzer, and 
Trakhtenbrot). On the positive side, the    
limited existing reverberation data for C\,{\sc iv} suggest that    
the $R$--$L$ relationship for C\,{\sc iv} has the same slope as    
\Hbeta\   
\cite[(Kaspi et al.\ 2007)]{Kaspi07}.  
Moreover, reverberation-based    
masses based on C\,{\sc iv} are consistent with those based on    
every other line (i.e., the virial product $\Delta V^2 R/G$ is
consistent for all the lines).
On the other hand, since C\,{\sc iv} is a    
resonance line, we often see absorption in its blue wing due to    
outflows, complicating accurate measurement of the line width.
Indeed, we sometimes see, particularly in the case of 
narrow-line Seyfert 1s (NLS1s),    
an extended blue wing of the emission line, again presumably    
due to outflowing gas, which  makes measuring the
line width problematic.
It seems likely, however, that we should be
able to calibrate out luminosity-dependent effects and
simply avoid spectra with strong absorption features
such as BALs.  
   
\subsection{Use of Scaling Relationships}   
The scaling relationships that lead to black-hole mass    
estimates must not be used blindly; indeed they should be used    
with great caution. We need to keep in mind that when we    
think we are measuring mass, we are really measuring   
\begin{equation}   
M_{\rm BH} \propto \Delta V^2 R \propto \Delta V^2    
L^{1/2}.   
\end{equation}   
Similarly, when we think we are measuring Eddington ratio,    
we are really measuring   
\begin{equation}   
\frac{L}{L_{\rm Edd}} \propto \frac{L}{M_{\rm BH}}    
\propto   
\frac{L}{\Delta V^2 L^{1/2}} \propto \frac{L^{1/2}}  
{\Delta V^2}.   
\end{equation}   
It is important to keep in mind that {\em any} correlations    
among mass, Eddington ratio, and luminosity must be eyed    
with great suspicion.   
   
\section {On the Possible Importance of  Radiation Pressure}   
\label{section:radpress}
Given a nominal bolometric correction, most AGNs are thought to  
be radiating at about $0.1 L_{\rm Edd}$ where  
$L_{\rm Edd}$ is the Eddington 
luminosity, the maximum luminosity at which the source is  
stable against disruption by radiation pressure.  
In the case of objects like  NLS1s, 
the luminosities may approach $L_{\rm Edd}$. It stands to reason  
that radiation pressure on the BLR gas might be an important  
dynamical force. In an elegant paper, 
\cite[Marconi et al.\ (2008)]{Marconi2008}  
argue that radiation pressure could be an important  
factor in the BLR dynamics. Radiation pressure, like gravity,   
is diluted geometrically (i.e., $\propto R^{-2}$),  
making it hard to distinguish between a high central  
mass plus radiation pressure and a lower central mass  
and no radiation pressure: failure to account for  
radiation pressure could thus lead us to underestimate  
the mass of the black hole. Marconi et al.\ argue that  
the relative importance of radiation pressure ultimately  
comes down to the inertia of the BLR gas clouds: if the  
clouds have sufficiently high column density  
($N_{\rm H} > 10^{23}$\,cm$^{-2}$),  
then the effect of radiation pressure on the gas dynamics  
is negligible. Interestingly, the most successful  
photoionization equilibrium calculations suggest  
column densities of this order. Marconi et al.\  
then modify equation (\ref{equation:virial})  to a form like 
\begin{equation} 
\label{equation:virialplusradiation}
M_{\rm BH} = f \Delta V^2 R/G + g L, 
\end{equation} 
where the first term on the right hand side is the 
mass based on the $R$--$L$ scaling relationship and  
the second term accounts for radiation pressure.  
Marconi et al.\ claim two successes with this  
formulation: first, the scatter between single-epoch  
mass estimates and reverberation measurements is  
reduced and the difference is no longer  
luminosity-dependent (their Figure 1), and second,  
NLS1s no longer fall below the normal  
\Msigma\ relationship. It is not clear,  
however, that anything physical can be inferred  
from this. The additional luminosity dependence  
of equation (\ref{equation:virialplusradiation})
can be attributed to the fact that the  
reverberation-based sample of AGNs is strongly  
Malmquist biased: the reverberation-mapped AGNs  
were selected because they are among the apparently 
brightest known AGNs, and there is a clear correlation  
between their black hole masses and luminosities. Thus the  
second term in equation (\ref{equation:virialplusradiation})
is bound to reduce scatter, regardless of  
whether or not radiation pressure is actually physically  
important. Moreover, the argument about NLS1s appears to be  
circular: if accounting for radiation pressure increases  
their black hole masses enough to place them on the  
\Msigma\  relationship, then their Eddington  
rates correspondingly drop down to those comparable to  
other AGNs, which then begs the question of the origin  
of the unusual properties of NLS1s. Also,  
\cite[Netzer (2009)]{Netzer2009}  
drew large samples of Type 1 and Type 2 AGNs from SDSS  
and showed that while their [O\,{\sc iii}] luminosity  
distributions and conventionally computed black hole masses  
have similar distributions, the radiation-pressure  
corrected mass distributions were markedly different,  
arguing against the importance of the radiation  
pressure term in equation (\ref{equation:virialplusradiation}).
\cite[Marconi et al.\ (2009)]{Marconi2009}
countered that Netzer's result is an evitable  
consequence of assuming a fixed column density for the  
BLR clouds; unfortunately, this argument exposes the  
fact that there is currently no accurate formulation  
for estimating black hole masses in the presence of radiation pressure.  
 
At the present time, whether or not radiation pressure  
needs to be accounted for in our mass calculations is  
not clear, though many of us continue to work on this issue.
We believe that NLS1s probably  
provide the best testing ground, and one obvious step  
is to further investigate the  
\Msigma\ relationship in these sources. Much of the previous  
work on NLS1s has necessitated using the [O\,{\sc iii}]  
narrow line widths as a surrogate for stellar  
velocity dispersions, and this is one area  
where we can make an improvement fairly easily by  
measuring stellar velocity dispersions in the near IR.

\section{Conclusions}
While great progress has been made in measuring the masses 
of the central black holes in
both active and quiescent galactic nuclei, there are still
significant uncertainties. My own guess is that masses
measured by all direct methods are probably uncertain by a factor of a few,
though this could be an underestimate for reverberation-based
masses if, for example, 
radiation pressure turns out to be important. Aside from
this possibility, the most significant 
unknown in reverberation-based
masses is the inclination of the system. This, of course,
is merely one aspect of a rather larger issue: it is also not
clear that we have identified an unbiased way to 
characterize the emission-line widths that enter into the
mass determination. The size of the BLR, at least as measured
with H$\beta$, seems to be remarkably well-characterized 
(though we could be fooling ourselves there, too) with
uncertainties as small as $ \Delta \log R \approx 0.1$\,dex.

\begin{acknowledgements}  
We are grateful for support of our work by
the National Science Foundation through grant
AST-0604066 and by NASA through grants
HST-GO-10833 and HST-GO-11661 from
Space Telescope Science Institute to
The Ohio State University.

\end{acknowledgements}

\end{document}